\documentclass[iop,useAMS]{emulateapj}

\usepackage{graphicx}
\usepackage[percent]{overpic}
\usepackage{multirow}
\usepackage{color}
\usepackage{rotating}

\newcommand{\msun}{${\rm M_{\sun}}$}

\def\ltsima{$\; \buildrel < \over \sim \;$}
\def\simlt{\lower.5ex\hbox{\ltsima}}
\def\gtsima{$\; \buildrel > \over \sim \;$}
\def\simgt{\lower.5ex\hbox{\gtsima}}
\def\kms{{\rm\,km\,s^{-1}}}
\def\kpc{{\rm\,kpc}}
\def\mpc{{\rm\,Mpc}}
\def\msun{{\rm\,M_\odot}}


\def\deg{^\circ}
\def\degg{\hbox{$\null^\circ$\hskip-3pt .}}

\def\Gyr{{\rm\,Gyr}}

\def\ltsima{$\; \buildrel < \over \sim \;$}
\def\gtsima{$\; \buildrel > \over \sim \;$}

\shorttitle{The halo of M31}
\shortauthors{Ibata et al.}

\begin{document}

\title{A thousand shadows of Andromeda:\\
rotating planes of satellites in the Millennium-II cosmological simulation}

\author{Rodrigo A. Ibata\altaffilmark{1}}
\author{Neil G. Ibata\altaffilmark{2}}
\author{Geraint F. Lewis\altaffilmark{3}}
\author{Nicolas F. Martin\altaffilmark{1,4}}
\author{Anthony Conn\altaffilmark{3}}
\author{Pascal Elahi\altaffilmark{3}}
\author{Veronica Arias\altaffilmark{3}}
\author{Nuwanthika Fernando\altaffilmark{3}}

\altaffiltext{1}{Observatoire astronomique de Strasbourg, Universit\'e de Strasbourg, CNRS, UMR 7550, 11 rue de l'Universit\'e, F-67000 Strasbourg, France; rodrigo.ibata@astro.unistra.fr}
\altaffiltext{2}{Lyc\'ee international des Pontonniers, 1 rue des Pontonniers, F-67000 Strasbourg, France}
\altaffiltext{3}{Institute of Astronomy, School of Physics A28, University of Sydney, NSW 2006, Australia}
\altaffiltext{4}{Max-Planck-Institut f\"ur Astronomie, K\"onigstuhl 17, D-69117 Heidelberg, Germany}

\begin{abstract}
In a recent contribution, \citet{2014MNRAS.438.2916B} investigated the incidence of planar alignments of satellite galaxies in the Millennium-II simulation, and concluded that vast thin planes of dwarf galaxies, similar to that observed in the Andromeda galaxy (M31), occur frequently by chance in $\Lambda$-Cold Dark Matter cosmology. However, their analysis did not capture the essential fact that the observed alignment is simultaneously radially extended, yet thin, and kinematically unusual. With the caveat that the Millennium-II simulation may not have sufficient mass resolution to identify confidently simulacra of low-luminosity dwarf galaxies, we re-examine that simulation for planar structures, using the same method as employed by Ibata et al. (2013) on the real M31 satellites. We find that 0.04\% of host galaxies display satellite alignments that are at least as extreme as the observations, when we consider their extent, thickness and number of members rotating in the same sense. We further investigate the angular momentum properties of the co-planar satellites, and find that the median of the specific angular momentum derived from the line of sight velocities in the real M31 structure ($1.3\times10^4 \kms \kpc$) is very high compared to systems drawn from the simulations. This analysis confirms that it is highly unlikely that the observed structure around the Andromeda galaxy is due to a chance occurrence. Interestingly, the few extreme systems that are similar to M31 arise from the accretion of a massive sub-halo with its own spatially-concentrated entourage of orphan satellites.
\end{abstract}


\section{Introduction}
\label{sec:Introduction}

The so-called $\Lambda$-CDM cosmology \citep{2011ApJS..192...18K}, in which the universe is dominated by dark energy and cold dark matter (CDM), accurately describes the large scale properties and evolution of the cosmos. On the scale of the halos of large galaxies, this model predicts copious CDM sub-structures, the most massive of which are identified with the dwarf satellite galaxies that are observed to inhabit such regions \citep{2011MNRAS.417.1260F}.
Numerical simulations that adopt this framework \citep{2008Springel,2010Cooper} generally reveal that the sub-halos that could host visible dwarf galaxies are distributed roughly spherically about the large galaxy.

Observationally however, the distribution of dwarf galaxies within the Local Group appears to be more complicated. \citet{1976MNRAS.174..695L} and \citet{1976RGOB..182..241K} first noted that several prominent dwarf galaxies are correlated with streams of \ion{H}{1} emission, and suggested that the outer halo globular clusters of the Milky Way may represent `ghostly' indicators of ancient accretions \citep{1995MNRAS.275..429L}. More recent analyses \citep{2007MNRAS.374.1125M,2008ApJ...680..287M,2009MNRAS.394.2223M,2012MNRAS.423.1109P} that include the faint dwarf galaxies detected during the past decade in the Sloan Digital Sky Survey (SDSS) support the earlier results, although concerns remain about the spatial selection biasses given the SDSS sky coverage. Interestingly, the satellites appear to be rotationally stabilized, orbiting within the plane defined by their spatial alignment \citep{2013MNRAS.435.2116P}. It has been claimed that the proposed planes of satellites are not predicted within $\Lambda$-CDM, and cannot simply represent a memory of past coherent accretion \citep{2005A&A...431..517K,2012MNRAS.423.1109P,2013MNRAS.435.1928P}, although other studies dispute this conclusion \citep{2005ApJ...629..219Z,2011MNRAS.413.3013L,2013MNRAS.429.1502W}. 

In a previous contribution we showed that our nearest large companion, the Andromeda galaxy, possesses an immense, kinematically coherent, thin plane of dwarf galaxies, representing $\sim 50\%$ of the total dwarf population of Andromeda (\citealt{2013Natur.493...62I}; see also \citealt{2013ApJ...766..120C}), confirming previous studies \citep{2006AJ....131.1405K,2007ApJ...670L...9M,2007MNRAS.374.1125M,2008ApJ...676L..17I,2009MNRAS.394.2223M} that had hinted at potential spatial correlations of dwarfs in M31.

\begin{figure*}
\begin{center}
\includegraphics[viewport= 75 155 560 640, clip, height=13cm]{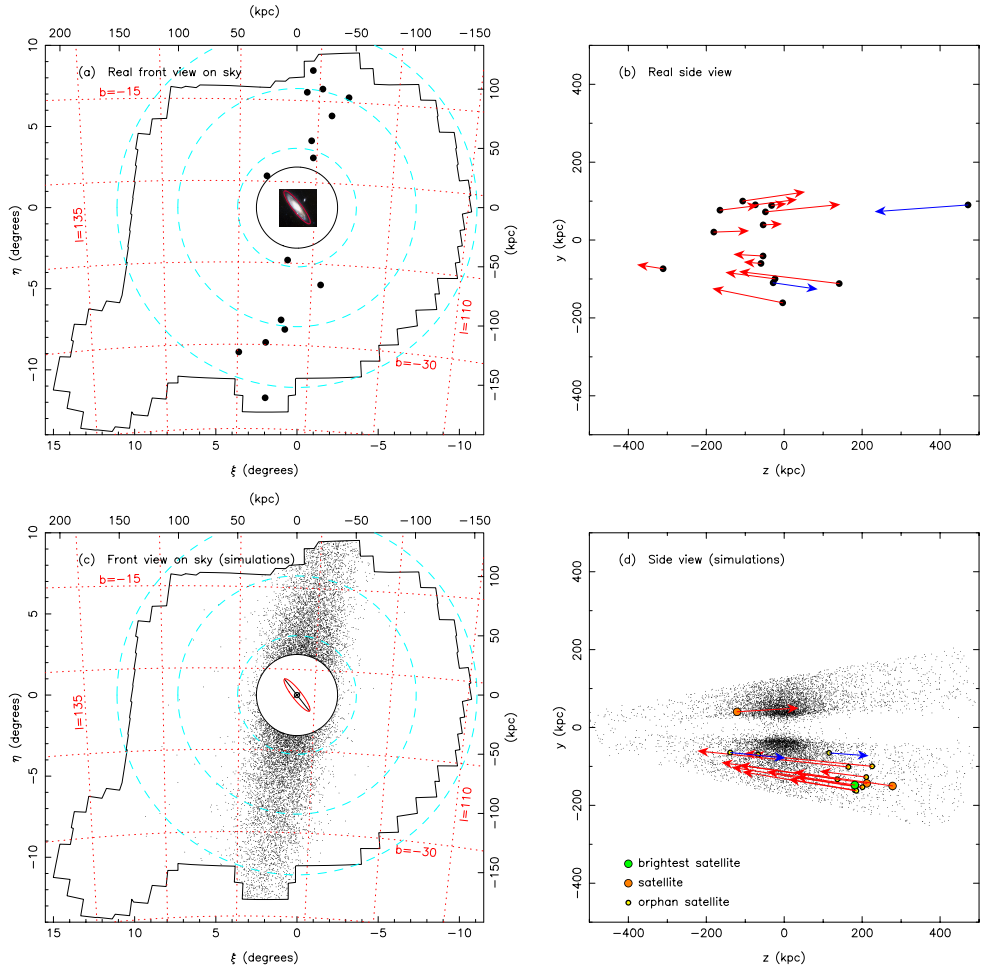}
\end{center}
\caption{Real and simulated alignments. Panel (a) shows the sky positions of the real sample of 15 satellite galaxies of M31 that display a planar alignment. (For objects at a distance of $780\kpc$, the top and right margins show the corresponding length scale, and dashed circles mark $50$, $100$ and $150\kpc$). The irregular polygon marks the outer limit of the Pan-Andromeda Archaeological Survey (PAndAS) \citep{2009Natur.461...66M}, while the inner (continuous) circle marks a $2\degg5$ region that was masked out by \citet{2013Natur.493...62I} to avoid incompleteness due to high stellar density. The side-on view in (b) shows the most likely positions of the satellites \citep{2012ApJ...758...11C}. The line of sight velocities of the satellites from \citet{2013ApJ...768..172C} are also displayed (a velocity of $1\kms$ has length $1\kpc$); the red arrows mark objects that share the same sense of rotation. The small dots in panels (c) and (d) show, respectively, the same information as in (a) and (b), but for all the satellites extracted from the Millennium-II simulation. In (d) we also overlay (large symbols) one of the two satellite systems that was more extreme than the observed M31 system. The green circle marks the position of the most massive sub-halo in that system, whose baryonic mass ($1.8\times10^{10}\msun$) significantly exceeds that of any satellite in the Local Group.}
\label{fig:observations}
\end{figure*}

A scenario that has been proposed to explain these alignments is that the present-day satellites are the remnants of tidal dwarf galaxies formed in ancient major mergers \citep{2012MNRAS.423.1109P,2013MNRAS.431.3543H}. This explanation is problematic, however, as it requires that the dwarfs are not dark matter dominated. It is in this context that \citet[][hereafter BB14]{2014MNRAS.438.2916B} recently analyzed the Millennium-II simulation \citep{2009MNRAS.398.1150B}, to investigate the incidence of satellite alignments in that large $10^6 h^{-3} \mpc^3$ volume of a $\Lambda$-CDM universe. They concluded that due to the spatial correlation between satellites in $\Lambda$-CDM, structures similar to that observed are relatively common, arising in approximately 2\% of the halos they investigated. The aim of the present paper is to examine the validity of the BB14 analysis, and to extend our earlier analysis to make better use of the accurate kinematic information available for the real satellites. In Section~\ref{sec:Sample_Selection}, we discuss how the samples were selected from the simulation. Section~\ref{sec:Results} presents the analysis and results, and conclusions are drawn in Section~\ref{sec:Conclusions}.

\section{Sample Selection}
\label{sec:Sample_Selection}

We select sub-halos from the Millennium-II simulation, using the \citet{2013MNRAS.428.1351G} catalog which was scaled to the WMAP year 7 analysis. The semi-analytic recipes that were applied to the dark matter simulation provide a wealth of predicted physical properties for galaxies that reside within the dark matter halos.

\begin{figure}
\begin{center}
\includegraphics[viewport= 65 80 560 740 , clip, width=8.5cm]{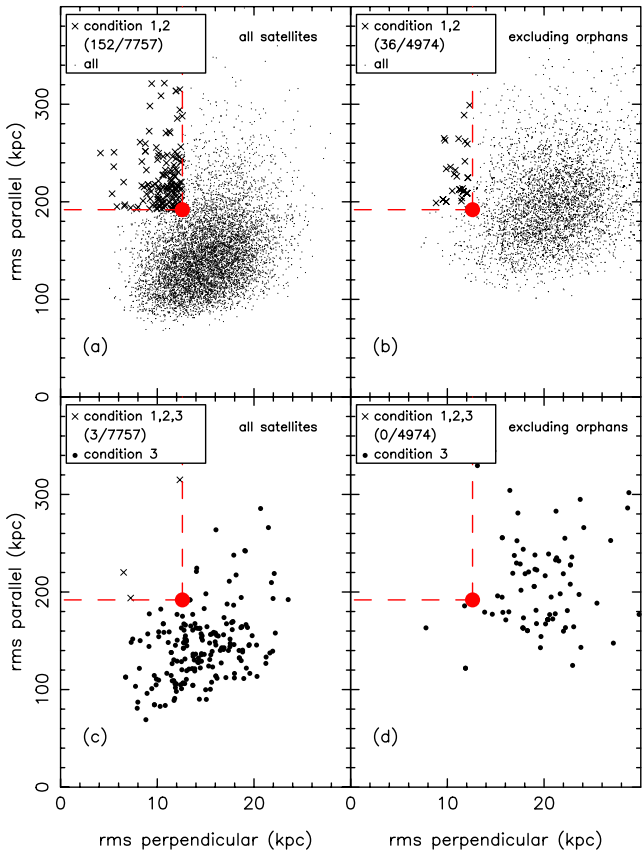}
\end{center}
\caption{The distribution of the dispersion within the plane of satellites (``rms parallel'') against the thickness of the plane (``rms perpendicular'') for the satellite systems. Panel (a) displays the samples that include orphan satellites. The large red dot marks the properties of the M31 satellite plane;  approximately 2\% of the Millennium-II systems (marked by crosses) are both at least as thin and as radially-extended as the M31 observations. However, the observed system also displays kinematic coherence. In panel (c) we display the subset of (a) that also has 13 or more satellites that possess the same sense of rotation. Now only 3 out of 7757 of the satellite systems (0.04\%), are both at least as thin and as radially-extended as the observations. These 3 systems correspond to 3 independent hosts. When orphan satellites are excluded (panels (b) and (d)), no satellite system in the Millennium-II simulation is found to be as extreme as M31.}
\label{fig:parameters}
\end{figure}

Since we are interested in selecting systems similar to M31, we follow BB14 and pick all halos in the redshift $z=0$ snapshot that have virial masses between $1.1\times10^{12}$ and $1.7\times10^{12}\msun$, mass-weighted age $<10\Gyr$, and whose neighbors within $500\kpc$ have a baryonic component less massive than $7\times10^{10}\msun$. (We define the baryonic mass to be the sum of the stellar and cold gas components). A total of 1141 such hosts are found. In addition to those BB14 selection criteria, we apply the following three criteria to select hosts that are reasonably similar to M31. We require that the hosts have baryonic masses in the range $2$--$20\times10^{10}\msun$ (this leaves 1024 galaxies), and we require that the group that the host galaxy resides in should have mass $<1.02\times10^{13}\msun$, a 95\% upper limit to the mass of the Local Group \citep{2008MNRAS.384.1459L} (leaving 885 galaxies). To ensure that the M31 analog, as in reality, is the dominant galaxy within the survey region, we finally reject hosts that have a companion within $500\kpc$ that possesses $>1/3$ of their baryonic mass. These cuts yield a sample of 679 host galaxies within the Millennium-II simulation.

Figure~\ref{fig:observations}a shows the observed distribution of the 15 satellites of M31 that are confined to a narrow plane. It can be immediately appreciated that the plane lies almost perfectly edge-on from our viewing point. The side view is displayed in Figure~\ref{fig:observations}b, using the most likely value of the line of sight distances measured by \citet{2012ApJ...758...11C}. The arrows mark the measured line of sight velocity \citep{2013ApJ...768..172C}, with the red arrows showing those satellites that have the same sense of rotation.

To reproduce the observational configuration, we place ourselves in the simulation at a random position $780\kpc$ distant from the host galaxy (equivalent to the Milky Way-M31 distance, \citealt{McConnachie:2005hn,2012ApJ...758...11C}), with a random camera angle. We initially select all the sub-halos within the Millennium-II simulation that lie within $500\kpc$ of the chosen host and that have stellar mass $>2.8\times 10^4 \msun$ (as proposed by BB14), to which we then apply the PAndAS spatial selection function\footnote{PAndAS was extended to the southeast to map M33; it is unlikely however that this peculiarity of the survey affects significantly our results.}. 

Within the PAndAS area shown in Figure~\ref{fig:observations}a, 27 satellites are known (beyond $2\degg5$), of which 15 are narrowly aligned in the thin plane discussed by \citet{2013Natur.493...62I}. To approximate this observed sample, we select in the simulation the brightest 27 satellites that are visible within the PAndAS field of view. If there are less than 27 satellites within this field of view, we discard the viewing position and draw a new one.

\begin{figure}
\begin{center}
\includegraphics[viewport= 65 80 560 740 , clip, width=8.5cm]{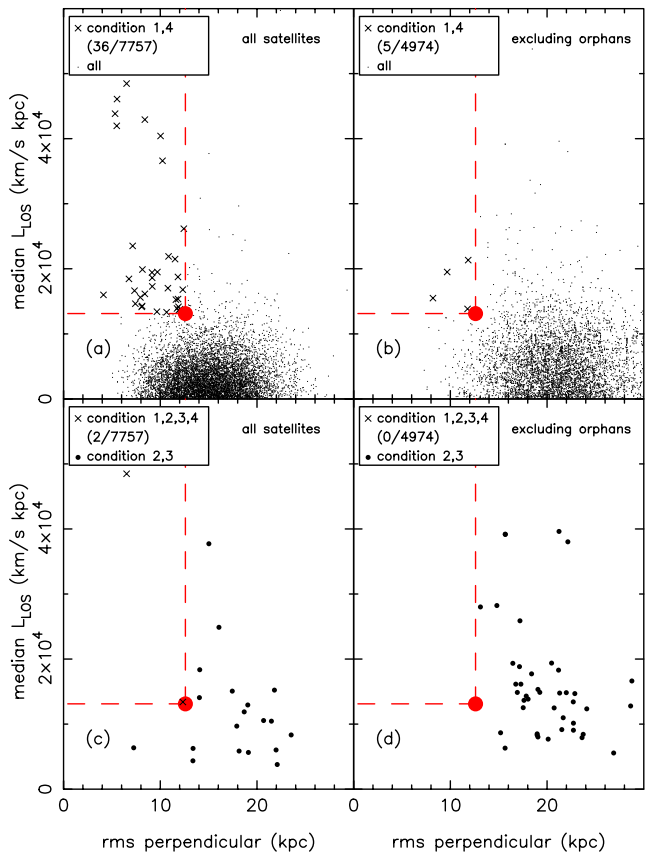}
\end{center}
\caption{The distribution of the median value of the angular momentum derived from line of sight velocities, $L_{\rm LOS}$, as a function of the rms thickness of satellite planes. Panel (a) includes orphan satellites, while (b) excludes them. The value for M31 is shown in red. The crosses in the upper panels mark those satellite systems that are both thinner and that possess higher median $L_{\rm LOS}$ than M31. In panels (c) and (d), we additionally impose the constraints on the radial extent and the coherent rotation. Only two satellite systems out of 7757 (0.03\%) in the Millennium-II sample remain after all constraints are applied (zero when no orphan galaxies are allowed).}
\label{fig:angular_momentum}
\end{figure}

Using exactly the same procedure as was applied to the real data in \citet{2013Natur.493...62I}, we calculate the pole of the plane that has the lowest perpendicular rms dispersion (i.e. the thinnest plane) for any sub-sample of 15 satellites. In a coordinate system where $x$ points east, $y$ points north and $z$ points along the line of sight vector to M31 (i.e. as shown in Figures~\ref{fig:observations}b and \ref{fig:observations}d), the pole of the observed structure points towards $(0.983,0.185,0.010)$. If the pole of the lowest rms plane, as seen from the randomly-chosen viewpoint, lies within $5\deg$ of the pole of the real plane, we store this host and its sample of satellites; otherwise a new viewing position is randomly drawn. If no sample is found after $10^6$ random draws of the viewing position, we discard the 
host. This procedure is repeated 12 times, so as to obtain (in the best cases) 12 different views of each of the 679 host galaxies. Cases where a new viewing position of the same host coincided within $15\deg$ to one previously calculated were discarded, as we considered such configurations to be ``repeats'' of the previous solutions\footnote{Ideally, each host galaxy would have been used only once, but since alignments are very rare, we were forced to boost the statistics. Fortunately, many hosts have a relatively large number ($\sim 100$) of satellites, meaning that different sub-samples are selected from different viewing positions. While our procedure gives rise to identical satellite sub-samples in 1.2\% of the random draws, the $>15\deg$ separation requirement still means that these provide different line of sight velocity projections (used in Figure~\ref{fig:angular_momentum}). In 76\% of the draws more than $1/3$ of the satellites are different. Hence the multiple views help build up better statistics.}. 

In Figures~\ref{fig:observations}c and \ref{fig:observations}d we overplot all the sub-samples of 15 satellites found around the host galaxies in the way described above. The samples have been rotated so that the viewpoint lies at $(x,y,z)=(0,0,-780)\kpc$. By construction, the random samples have the same orientation as the data. Our main motivation to force the alignments to be edge-on was to be able to compare the radial velocity information on the satellites directly to the simulations, but this also automatically corrects for the inclination-dependent detection bias in PAndAS  \citep{2013ApJ...766..120C}, and it also avoids the issue that rotation of a structure aligned perpendicular to the line of sight would not have been detected. Nevertheless, tests show that forcing this orientation does not substantially alter the results below.

\section{Analysis and discussion}
\label{sec:Results}

The semi-analytic modeling of \citet{2013MNRAS.428.1351G} differentiates normal galaxies from ``orphans''; the latter being systems whose parent subhalo is no longer resolved. It is possible that many of these orphans are tidally disrupted, and that they are hence not directly comparable to the observed dwarf galaxies. Despite these concerns, we will first assume that the orphan satellites represent structures that can be reliably compared to real dwarf galaxies (this is an assumption that BB14 considered). However, we noticed that in the Millennium-II simulation, many of the orphans are spatially clustered around massive sub-halos. We decided to reject all the orphan satellites that were next to a massive, non-orphan, satellite, closer than twice the radius of the stellar disk of the non-orphan satellite. This is justified on the grounds that in reality it is unlikely that such objects would have been identified (they would be very hard to distinguish from their companion). Additionally, we reject orphan satellites that are both closer than $20\kpc$ (which corresponds approximately to the line of sight distance uncertainty for the best measured dwarf galaxies) and that have space velocities $<50\kms$ from a non-orphan satellite, since observationally we would consider such objects to be tidal debris.

This is the sample that was displayed previously in Figures~\ref{fig:observations}c and \ref{fig:observations}d. One notices immediately from a visual comparison of panels (b) and (d) that the sample of simulated satellites that includes orphans is significantly more centrally-concentrated than the real satellites, so clearly the extent of the plane is an important property to attempt to reproduce, in addition to its thinness. BB14 decided to parametrize the radial extent via the rms dispersion within the plane (``rms parallel''), which we calculate as the sum of the squares of the distances in the plane {\it from the host galaxy}. In Figure~\ref{fig:parameters}a we show this parameter plotted against the plane thickness (``rms perpendicular''). 

Let us define ``condition 1'' to be the requirement that the simulated systems should be thinner than the observed structure ($rms_{\rm per} <12.6\kpc$), ``condition 2'' that they should be spatially more extended than the observations ($rms_{\rm par} >191.9\kpc$), and ``condition 3'' that 13 or more out of the subsamples of 15 satellites share the same sense of rotation. We find that 2\% of the systems $(152/7757)$ conform to conditions 1 and 2 (crosses in panel (a)), but only 3 out of 7757 simulated systems (0.04\%) conform to conditions 1,2 and 3 (crosses in panel (c)).

If instead we discard all the orphan satellites, we obtain the parameter distributions displayed in panels (b) and (d) of Figure~\ref{fig:parameters}. In this case not a single system within the Millennium-II simulation out of a sample of 4974 is more extreme than the M31 satellite structure (upper limit of $0.02$\%). Note that we have been very conservative in the selection, and have not counted any system in which we could not find 27 satellites of which 15 are in an edge-on configuration (which is why the number of systems in the right panels of Figure~\ref{fig:parameters} is lower than in the left panels).

An interesting and useful additional view of the parameter space of the properties of the satellite structure is given in Figure~\ref{fig:angular_momentum}. Since we deliberately enforced the simulated alignments to be seen with the same orientation as the observations, we can directly compare the line of sight velocities of the satellites with the real measurements. In reality we do not have access to the proper motions of the satellites, neither do we have reliable total mass estimates, so we cannot measure their total angular momentum. However, it is possible to calculate the minimum specific angular momentum, $L_{\rm LOS}$, simply by using the line of sight velocities and taking the unknown transverse motions to be zero. Exactly the same statistic can be calculated for the simulated satellite galaxies, so the comparison should be very robust. 

In many of the simulated systems, there are a few satellites with extreme values of $L_{\rm LOS}$. To avoid being dominated by such outliers, we chose to adopt (the absolute value of) the median of $L_{\rm LOS}$ as our statistic. This parameter is displayed along with the rms thickness in Figure~\ref{fig:angular_momentum}, for which we define a further ``condition 4'': that the simulated systems should have median $L_{\rm LOS} >1.3\times 10^4 \kms \kpc$, the value of the $L_{\rm LOS}$ in the real data\footnote{An extended structure with bulk transverse velocity will show projections of that velocity onto the line of sight that increase outwards from its center.  We may estimate the magnitude of this effect by assuming that the \citet{2012ApJ...753....8V} proper motion measurement of M31 approximates the bulk motion of the observed planar system. With this, we derive a slightly lower value of $L_{\rm LOS} =1.2\times 10^4 \kms \kpc$, which implies that bulk motion is not likely to be important for the present discussion.}. The bottom panels show the result of applying conditions 1 to 4 to the simulated systems. If we discount orphan satellites, there are no structures in the Millennium-II simulations that compare to what is seen in M31, but if we allow orphans, 2 out of 7757 systems (0.03\%) are more extreme in their thinness, radial extent and kinematic properties than the real structure around Andromeda.

It is interesting to inspect these two cases closely. Both systems contain very massive satellites, with virial masses of $2.3\times10^{11}\msun$ and $5.8\times10^{11}\msun$, representing mass fractions of 1$:$5 and 1$:$3, and with baryonic masses of $1.1\times10^{10}\msun$ and $1.8\times10^{10}\msun$ (baryonic mass fractions of 1$:$7 and 1$:$4), respectively. Furthermore, each of these halos has an accompanying compact entourage of orphan companions (unlike the spread-out real system), as can be seen in Figure~\ref{fig:observations}d, where we plot the positions and velocities of one of these systems.  Finally, in both cases, the massive companion is approaching the host for the first time. This suggests the possibility that the observed structure in M31 is due to the accretion of such a system, and indeed the baryonic masses listed above seem to be in reasonable agreement with that of M33, which has a stellar mass of $3$--$6\times10^9\msun$ and a total gas mass of $3.2\times10^9\msun$ \citep{2003MNRAS.342..199C}. It should nevertheless be kept in mind that these accretions of a massive satellite galaxy with a tight sub-system of their own are {\it extremely rare} events in the Millennium-II simulations. Note also that in reality M33 does not partake in the observed planar alignment.

What is the origin of the discrepancy between the conclusions of our work and BB14? The limitation with their study is that when they examine the kinematics of the alignments, they only use one of the structural parameters ($rms_{\rm per}$), not both ($rms_{\rm per}$ and $rms_{\rm par}$)\footnote{To make the issue perfectly clear, consider measuring the incidence of animals that have stripes {\it and} paws {\it and} are nocturnal. Clearly, selecting on only two of these three properties will yield a larger (and incorrect) sample of such animals, giving a falsely optimistic measurement of how common they are.}. Once this error is corrected, our results and theirs are surprisingly consistent (i.e., our more conservative selection of hosts and the proper PAndAS spatial selection function applied to the satellites do not substantially affect the result): both studies measure a $\sim0.04$\% occurrence rate\footnote{BB14 measured that 2\% of satellites obey both condition 1 {\it and} condition 2, and also that 2\% obey both condition 1 {\it and} condition 3. This implies a $\sim0.04$\% joint probability only if conditions 2 and 3 are independent.}.

\section{Conclusions}
\label{sec:Conclusions}

We have carefully reanalyzed the large-scale ``Millennium-II'' $\Lambda$-CDM simulation, searching for alignments of satellites similar to that observed around the Andromeda galaxy. We consider M31-like host galaxies in a range of a full decade in stellar mass around the M31 value and require that they reside in parent dark matter halos that are no more massive than plausible values for the Local Group.

By applying the PAndAS spatial selection function we derive views of planes of 15 satellites that are comparable to the observed configuration. We analyzed the perpendicular thinness, radial extent, coherent kinematics and angular momentum properties of the simulated samples together, and found cases similar to the observed planar structure to be extremely rare, occurring in only 0.03--0.04\% of the samples. 

This shows that the observed alignment discovered by \citet{2013Natur.493...62I} is surprising in $\Lambda$-CDM, if one assumes that the Millennium-II simulation has sufficient resolution to reliably detect counterparts of the observed satellites. Nevertheless, the extreme rarity of analogs to the M31 system in the Millennium-II simulation does not necessarily preclude the possibility that planar structures may exist around different types of hosts in $\Lambda$-CDM simulations. By relaxing our search criteria that are specific to M31, such as the adopted virial and baryonic mass constraints, and the requirement that there be no nearby large neighbor, the number of candidate systems increases from 679 to over 2000 for virial masses in the range $1$--$5\times10^{12}\msun$. While a thorough analysis of such simulated satellite systems is beyond the scope of the present paper, we believe that it will be of great importance to ascertain the incidence of satellite alignments in nature for galaxies beyond the Local Group.

\acknowledgments

The Millennium-II Simulation databases used in this paper were constructed as part of the activities of the German Astrophysical Virtual Observatory (GAVO).


\begin{thebibliography}{69}
\expandafter\ifx\csname natexlab\endcsname\relax\def\natexlab#1{#1}\fi


\bibitem[Bahl 
\& Baumgardt(2014)]{2014MNRAS.438.2916B} Bahl, H., \& Baumgardt, H.\ 2014, \mnras, 438, 2916 

\bibitem[Boylan-Kolchin et al.(2009)]{2009MNRAS.398.1150B} Boylan-Kolchin, 
M., Springel, V., White, S.~D.~M., Jenkins, A., 
\& Lemson, G.\ 2009, \mnras, 398, 1150 

\bibitem[{Conn {et~al.}(2012)Conn, Ibata, Lewis, Parker, Zucker, Martin,
  McConnachie, Irwin, \& {et al.}}]{2012ApJ...758...11C}
Conn, A.~R., {et~al.} 2012, ApJ, 758, 11

\bibitem[Conn et al.(2013)]{2013ApJ...766..120C} Conn, A.~R., Lewis, G.~F., 
Ibata, R.~A., et al.\ 2013, \apj, 766, 120 

\bibitem[Collins et al.(2013)]{2013ApJ...768..172C} Collins, M.~L.~M., 
Chapman, S.~C., Rich, R.~M., et al.\ 2013, \apj, 768, 172 

\bibitem[{{Cooper} {et~al.}(2010){Cooper}, {Cole}, {Frenk}, {White}, {Helly},
  {Benson}, {De Lucia}, {Helmi}, {Jenkins}, {Navarro}, {Springel}, \&
  {Wang}}]{2010Cooper}
{Cooper}, A.~P., {et~al.} 2010, \mnras, 406, 744

\bibitem[Corbelli(2003)]{2003MNRAS.342..199C} Corbelli, E.\ 2003, \mnras, 
342, 199 

\bibitem[{{Font} {et~al.}(2011){Font}, {Benson}, {Bower}, {Frenk}, {Cooper},
  {De Lucia}, {Helly}, {Helmi}, {Li}, {McCarthy}, {Navarro}, {Springel},
  {Starkenburg}, {Wang}, \& {White}}]{2011MNRAS.417.1260F}
{Font}, A.~S., {et~al.} 2011, \mnras, 417, 1260

\bibitem[Guo et al.(2013)]{2013MNRAS.428.1351G} 
Guo, Q., White, S., Angulo, R.~E., et al.\ 2013, \mnras, 428, 1351 

\bibitem[Hammer et al.(2013)]{2013MNRAS.431.3543H} Hammer, F., Yang, Y., 
Fouquet, S., et al.\ 2013, \mnras, 431, 3543 

\bibitem[{Ibata {et~al.}(2013)Ibata, Lewis, Conn, Irwin,
  McConnachie, Chapman, Collins, Fardal, \& {et al.}}]{2013Natur.493...62I}
Ibata, R.~A., {et~al.} 2013{\natexlab{b}}, Nature, 493, 62

\bibitem[{{Irwin} {et~al.}(2008){Irwin}, {Ferguson}, {Huxor}, {Tanvir},
  {Ibata}, \& {Lewis}}]{2008ApJ...676L..17I}
{Irwin}, M.~J., {Ferguson}, A.~M.~N., {Huxor}, A.~P., {Tanvir}, N.~R., {Ibata},
  R.~A., \& {Lewis}, G.~F. 2008, \apjl, 676, L17

\bibitem[{{Koch} \& {Grebel}(2006)}]{2006AJ....131.1405K}
{Koch}, A., \& {Grebel}, E.~K. 2006, \aj, 131, 1405

\bibitem[{{Komatsu} {et~al.}(2011){Komatsu}, {Smith}, {Dunkley}, {Bennett},
  {Gold}, {Hinshaw}, {Jarosik}, {Larson}, {Nolta}, {Page}, {Spergel},
  {Halpern}, {Hill}, {Kogut}, {Limon}, {Meyer}, {Odegard}, {Tucker}, {Weiland},
  {Wollack}, \& {Wright}}]{2011ApJS..192...18K}
{Komatsu}, E., {et~al.} 2011, \apjs, 192, 18

\bibitem[Kroupa et 
al.(2005)]{2005A&A...431..517K} Kroupa, P., Theis, C., \& Boily, C.~M.\ 2005, \aap, 431, 517 

\bibitem[Kunkel 
\& Demers(1976)]{1976RGOB..182..241K} Kunkel, W.~E., \& Demers, S.\ 1976, The Galaxy and the Local Group, 182, 241 

\bibitem[Li \& White(2008)]{2008MNRAS.384.1459L} Li, Y.-S., \& White, S.~D.~M.\ 2008, \mnras, 384, 1459 

\bibitem[{{Lovell} {et~al.}(2011){Lovell}, {Eke}, {Frenk}, \&
  {Jenkins}}]{2011MNRAS.413.3013L}
{Lovell}, M.~R., {Eke}, V.~R., {Frenk}, C.~S., \& {Jenkins}, A. 2011, \mnras,
  413, 3013

\bibitem[{{Lynden-Bell}(1976)}]{1976MNRAS.174..695L}
{Lynden-Bell}, D. 1976, \mnras, 174, 695

\bibitem[{{Lynden-Bell} \& {Lynden-Bell}(1995)}]{1995MNRAS.275..429L}
{Lynden-Bell}, D., \& {Lynden-Bell}, R.~M. 1995, \mnras, 275, 429

\bibitem[{{Majewski} {et~al.}(2007){Majewski}, {Beaton}, {Patterson},
  {Kalirai}, {Geha}, {Mu{\~n}oz}, {Seigar}, {Guhathakurta}, {Gilbert}, {Rich},
  {Bullock}, \& {Reitzel}}]{2007ApJ...670L...9M}
{Majewski}, S.~R., {et~al.} 2007, \apjl, 670, L9

\bibitem[van der Marel et al.(2012)]{2012ApJ...753....8V} van der Marel, 
R.~P., Fardal, M., Besla, G., et al.\ 2012, \apj, 753, 8 

\bibitem[{McConnachie {et~al.}(2005)McConnachie, Irwin, Ferguson, Ibata, Lewis,
  \& Tanvir}]{McConnachie:2005hn}
McConnachie, A.~W., Irwin, M.~J., Ferguson, A. M.~N., Ibata, R.~A., Lewis,
  G.~F., \& Tanvir, N. 2005, MNRAS, 356, 979

\bibitem[{McConnachie {et~al.}(2009)McConnachie, Irwin, Ibata, Dubinski,
  Widrow, Martin, C{\^o}t{\'e}, Dotter, \& {et al.}}]{2009Natur.461...66M}
McConnachie, A.~W., {et~al.} 2009, Nature, 461, 66

\bibitem[{{Metz} {et~al.}(2007){Metz}, {Kroupa}, \&
  {Jerjen}}]{2007MNRAS.374.1125M}
{Metz}, M., {Kroupa}, P., \& {Jerjen}, H. 2007, \mnras, 374, 1125

\bibitem[{{Metz} {et~al.}(2008){Metz}, {Kroupa}, \&
  {Libeskind}}]{2008ApJ...680..287M}
{Metz}, M., {Kroupa}, P., \& {Libeskind}, N.~I. 2008, \apj, 680, 287

\bibitem[{{Metz} {et~al.}(2009){Metz}, {Kroupa}, \&
  {Jerjen}}]{2009MNRAS.394.2223M}
---. 2009, \mnras, 394, 2223

\bibitem[{{Pawlowski} {et~al.}(2012){Pawlowski}, {Pflamm-Altenburg}, \&
  {Kroupa}}]{2012MNRAS.423.1109P}
{Pawlowski}, M.~S., {Pflamm-Altenburg}, J., \& {Kroupa}, P. 2012, \mnras, 423,
  1109

\bibitem[Pawlowski et al.(2013)]{2013MNRAS.435.1928P} Pawlowski, M.~S., 
Kroupa, P., \& Jerjen, H.\ 2013, \mnras, 435, 1928 

\bibitem[Pawlowski 
\& Kroupa(2013)]{2013MNRAS.435.2116P} Pawlowski, M.~S., \& Kroupa, P.\ 2013, \mnras, 435, 2116 

\bibitem[{{Springel} {et~al.}(2008){Springel}, {Wang}, {Vogelsberger},
  {Ludlow}, {Jenkins}, {Helmi}, {Navarro}, {Frenk}, \& {White}}]{2008Springel}
{Springel}, V., {et~al.} 2008, \mnras, 391, 1685

\bibitem[Wang et al.(2013)]{2013MNRAS.429.1502W} Wang, J., Frenk, C.~S., 
\& Cooper, A.~P.\ 2013, \mnras, 429, 1502 

\bibitem[{{Zentner} {et~al.}(2005){Zentner}, {Kravtsov}, {Gnedin}, \&
  {Klypin}}]{2005ApJ...629..219Z}
{Zentner}, A.~R., {Kravtsov}, A.~V., {Gnedin}, O.~Y., \& {Klypin}, A.~A. 2005,
  \apj, 629, 219


\end{thebibliography}

\end{document}